# Optical Properties of ITO/ZnO Schottky Diode with Enhanced UV Photoresponse

**Hsin-Yen Lee and Ming-Yau Chern**

*Department of Physics, National Taiwan University, Taipei, 10617, Taiwan (R.O.C.)*

Photoluminescence (PL) spectra of zinc oxide (ZnO) samples with different hydrogen peroxide ($H_2O_2$) treatment durations were measured to examine several point defects on the surface of the films. These results suggest successful oxidation through the reaction between oxygen radicals dissociated from $H_2O_2$ and the ZnO surface. To further confirm the defect induced gain mechanism, we fabricate highly transparent indium–tin-oxide (ITO)/ZnO Schottky diodes, and measure the key diode characteristics. Photocurrents are measured under different wavelengths, and possible explanations of the high optical gain within the ultraviolet (UV) region are provided.

PACS number: 72.40.+w, 81.05.Dz, 81.40.Tv, 81.40.-z

Keywords: ZnO, ITO, Photoluminescence, Photocurrent

Email: mychern@phys.ntu.edu.tw

Fax: +886-2-2363-9984

# I. INTRODUCTION

ZnO, an n-type semiconductor with a direct wide bandgap of 3.36 eV, has been prepared by scientists for various applications. Especially useful ones are the devices operating in the UV region [1-6]. Despite its popularity, scientists continue to search for the origin of its n-type conductivity in ZnO, both theoretically and experimentally. There are six native point defects which have been identified: zinc vacancies ($V_{Zn}$), oxygen vacancies ($V_O$), zinc interstitials ($Zn_i$), oxygen interstitials ($O_i$), zinc antisites ($Zn_O$), and oxygen antisites ($O_{Zn}$). Among these defects, $Zn_i$, $Zn_O$, and $V_O$ are considered to be native donors, whereas $O_i$, $O_{Zn}$, and $V_{Zn}$ are believed to be native acceptors [7]. Experimentalists generally believed that these intrinsic native donors, $V_O$ and $Zn_i$ in particular, cause the natural n-type characteristic of ZnO [8]. Theoretical physicists, on the other hand, offered different opinions. For example, Zhang *et al.* [9] showed that, by using the density-functional theory with the local density approximation (DFT-LDA), when grown under Zn-rich condition or O-rich condition, $Zn_i$, which is a shallow donor, has low formation enthalpy. On the other hand, native defects that could compensate the n-type doping ($O_i$, $V_{Zn}$) have high formation enthalpies. Janotti *et al.* [10] showed that $V_O$ is actually a deep donor at approximately 1 eV below the conduction band and cannot introduce enough carriers because of its large ionization energy. More recently, hydrogen (H) has become a popular choice to be responsible for the n-type conductivity of ZnO [11, 12]. Theoretical calculations showed that H is a shallow donor in ZnO [13], and experimental results indicated that the conductivity of ZnO increases in $H_2$-ambient conditions [14]. All of the above studies showed that the origin of n-type conductivity of ZnO is still controversial, and therefore more experimental data and more advanced techniques [15-17] are still needed. In this work, $H_2O_2$-treated highly c-axis-oriented ZnO samples grown by using laser deposition method were examined by room temperature (RT) PL to experimentally discover the nature of ZnO n-type conductivity and the best film-growing environment. Furthermore, to confirm the defects induced gain mechanism of ZnO diodes, vertical structured Schottky diodes with ITO as the metallic contact were grown, and some important diode parameters, including the ideality factor ($n$) and Schottky barrier height (SBH), were fitted individually. Similar device was fabricated previously by using chemical vapor deposition (CVD) grown ZnO nanowires [18], and the measured $n = 1.82$ and SBH = 0.89 eV prove its feasibility and high quality. Finally, the responsivity ($R$) at a given wavelength, and the explanation of the high gain effect is also given.

## II. EXPERIMENTS AND DISCUSSION

The ZnO samples used in this research were grown by a laser evaporation process using a continuous wave (CW) carbon dioxide ($CO_2$) laser. The laser model was a Coherent GEM-100L RF-excited $CO_2$ laser with a wavelength of 10.6 µm and a maximum output power of 100 W. The $CO_2$ laser was modulated by a function generator to produce 5 Hz pulses with energy of 4 J/pulse. The incidence angle of the $CO_2$ laser was 30° to the normal of both the ZnO and ITO targets. The targets vaporized after the absorption of the laser energy, and then condensed on the substrate directly facing the target at a distance of 35 mm. All of the ZnO samples were grown at 600 °C without post-annealing, and all of the metal layer ITOs deposited on ZnO were grown at RT. All of the substrates employed in this research were ITO coated glass (ITO-glass) and $6 \times 8$ mm$^2$ in size. Prior to deposition, the substrates were ultrasonically degreased in acetone and alcohol for 10 min per stage to remove contaminants, rinsed in deionized water, and then dried under nitrogen flow. All of the ZnO samples underwent a post-annealing process at their growth temperature for 30 min followed by a 70 °C $H_2O_2$ treatment for 20 min. An $H_2O_2$ solution (35 %, 70 °C) was used for surface treatment under four durations (i.e., 0, 600, 1200, and 1800 s).

PL data were collected by using a He-Cd laser ($\lambda = 325$ nm) at RT. Photons are detected by a detector connected by an optical fiber into a USB2000+VIS-NIR miniature spectrometers ranging from 350 nm to 1000 nm. Photocurrents were measured by a monochromatic Xe-lamp and an Acton Research Corporation SpectraDrive spectrometer, and the incident light power data acquisition was recorded by a Newport Dual-Channel Power Meter Model 2832-C power meter. The electrical conductivity of the ZnO films was estimated by the Van der Pauw techniques. The setup consists of a Keithley 7065 Hall effect card, a 220 current source, and a 6485 picoammeter.

The PL behavior of the ZnO thin films grown at 600 °C under different $H_2O_2$ treatment durations is shown in Fig. 1. Each PL spectrum is vertically shifted for clarity. The PL spectra exhibit two peaks: a near-band-edge (NBE) emission at 377.6 nm (which corresponds to 3.28 eV) and a deep-level (DL) emission in the visible region (yellow–green) centered at approximately 580 nm (2.14 eV). NBE emission is caused by a band-to-band transition, where a free electron in the conduction band recombines with a free hole in the valence band, and a transition of free excitons, where recombination of the electron and hole of a moving exciton causes the emission of a photon with energy slightly less than the bandgap energy. DL emission is generally believed by researchers to be caused by native defects within the bandgap of ZnO: yellow emission is assigned to $O_i$ and green emission may originate from surface defects, such as $V_o$ [19]. Results showed that as the treatment duration increases, the peak of the NBE emission at 377.6 nm increases and the intensity of the broad-band DL emission decreases. DL emissions not only is decreased but also becomes broadened as the treatment time increased. The ratio of NBE to DL emissions is estimated to rise from 6.97 (0 s) to 7.51 (600 s), 13.70 (1200 s), and 20.67 (1800 s). The increasing NBE emission peak intensity may be related to the size effect of the nanostructure of the film surface [20]. The decrease in DL emission suggests a reduction in $V_o$ and the broadened yellow emission indicates oxidation. This result validates the effects of $H_2O_2$, where oxygen radicals dissociated from the $H_2O_2$ react with the ZnO surface, remove the OH contamination layer, fill the vacancies, produce possible interstitials, and form ZnO and $ZnO_2$.

The advantage of $H_2O_2$ processing is that it increases the effective SBH and decreases the leakage current, which leads to better diode characteristics, as the diode turn from Ohmic to Schottky after the treatment. We note that the increased $O_i$ on the surface will lead to high gain. In order to examine this effect, several ITO/ZnO Schottky diodes were fabricated, with all of them treated with $H_2O_2$ for 1800 sec. Plot of the experimental SBH versus *n* is shown in Fig. 2, and a current-voltage (*I-V*) characteristic curve with fitting results are shown in the inset. Average carrier concentration of the films without $H_2O_2$ treatment is $2.35\times10^{17}$ cm$^{-3}$, and is $2.26\times10^{17}$ cm$^{-3}$ for films with $H_2O_2$ treatment. Mobility for films without $H_2O_2$ treatment ranges from 22.07 – 24.30 cm$^2$/V-sec without, and ranges from 19.45 – 23.35 cm$^2$/V-sec for films with $H_2O_2$ treatment. Most of the diodes have *n* and SBHs ranging from 1 – 10 and 0.6 – 0.8 eV, respectively. It should be noted that although these diodes were grown at the same temperature and have the same treatment duration, other parameters, such as oxygen pressure and the heating process, were somehow different. This explains the difference in the diode characteristics, especially for ones with very large *n*. A downward trend in this figure can be observed, and this linear relationship is mostly attributed to lateral inhomogeneous interfaces, as theoretical calculations and numerical simulations are studied [21].

Optical data were gathered from a high quality (*n* = 2.45, SBH = 0.82 eV under dark conditions) diode. Figure 3 shows *R* in the wavelength region from 280 nm to 480 nm under a reversed bias voltage of 0.5 V. Important data extracted from the *I-V* characteristics of the corresponding diode in dark under $\lambda$ = 390 nm illumination are given in Table 1. Upon UV illumination (from 280 nm to 380 nm), large measured photocurrent $I_p$ were obtained and a maximum $2.98\times10^{-4}$ A was realized at 390 nm. Over 400 nm, the measured $I_p$ dropped significantly to the order of $10^{-6}$ A. Spectral responsivity *R*, which is defined as $I_p$ per incident optical power at a given wavelength (see the inset of Fig. 3), reached a maximum at 300 nm, with a measured value of 90.8 A/W, and remained in the same order between 70 and 90 A/W before dropping off by about two orders of magnitude to 0.91 A/W at 400 nm. This high *R* can be attributed to the high gain of the diodes, which is commonly reported in GaN and ZnO-based devices [22]. One of the possible explanations may be the enhanced electron tunneling effect, which states that the thermionic-field emission dominates in diodes under above-bandgap illumination ($\lambda \leq 390$ nm) instead of thermionic emission, which dominates in diodes under dark conditions [23]. The tunnel current becomes significant and the diode exhibits Ohmic behavior under illumination. A second explanation for the phenomenon observed is that the high gain corresponds to an enhanced secondary photoresponse caused by minority carrier trapping at the metal–semiconductor interface, resulting in a Schottky barrier reduction under illumination [24]. Due to the low leakage current and low fitted $n$ under both dark and illuminated conditions, the enhanced electron tunneling can hardly be the reason for such a high gain at only a 0.5 V reversed bias voltage. Thus we conclude that in our case, high gain is caused by the latter explanation.

In order to support this conclusion, we offer a quantitative analysis to compare our experimental data with secondary photoresponse model. Generally speaking, the major current transport process of the Schottky diode can be explained by the thermionic-emission theory [25]. Together with other possible processes, such as tunneling, recombination, and diffusion of the electrons and holes, the total current density can be expressed as [26]

$$I = A^*T^2\exp\left(-\frac{\phi_{Bn}}{kT}\right)\left[\exp\left(\frac{qV}{nkT}\right) - 1\right], \quad (1)$$

where $q$ is the electron charge, $V$ is the applied voltage, $k$ is the Boltzmann constant, $T$ is the absolute temperature ($\cong 300$K), $A^*$ is the effective Richardson constant ($\cong 32$ A cm$^{-2}$ K$^{-2}$), and $\phi_{Bn}$ is the SBH, which can be obtained through the saturation current $I_S$ by

$$I_S = A^*T^2\exp\left(-\frac{\phi_{Bn}}{kT}\right), \quad (2)$$

From values extracted from the *I-V* measurements, we have $\phi_{Bn} = 1.42$ eV for diode in the dark and $\phi_{Bn} = 1.18$ eV under illumination, which give us the reduction of SBH $\Delta\phi_{Bn} = 0.24$ eV. According to the SBH reduced gain mechanism, $R$ can be calculated as

$$R = \frac{[\exp(\Delta\phi_{Bn}/kT)-1]I_{dark}-I_\lambda}{W} \quad (3)$$

In the above equation, $I_\lambda$ is the measured photocurrent at $\lambda$ and $W$ is the light intensity. This formula suggests that the photoresponse comes from a primary current $I_\lambda$ by carriers generated in the depletion region, and a secondary current generated by the reduction of SBH. Consider $\lambda = 390$ nm near the absorption edge, $I_{dark} = 4.30 \times 10^{-5}$ A, $I_\lambda = 2.98 \times 10^{-4}$ A, $\Delta\phi_{Bn} = 0.24$ eV and $W = 8.45 \times 10^{-3}$ W/m², we have $R = 54.7$, which is in good agreement with measured data ($R = 72.9$). Note that this value is only a rough approximation, as the calculated $R$ can be significantly influenced by the reduction of SBH $\Delta\phi_{Bn}$. Also, the uncertainty in the assumed Richardson constant $A^*$ can affect the estimation. Nevertheless, values estimated from *I-V* characteristics and predicted theoretically show that a major contribution of the high gain is indeed caused by the secondary photoresponse, and the lower the reduced SBH, the higher the responsivity $R$.

## III. CONCLUSION

In conclusion, this letter presents the relationship between ZnO surface point defects and the durations of H2O2 treatment. Reduced oxygen vacancies and gradual oxidation are obtained from the decrease in DL emission and broadened yellow emission. The reduced SBH caused by a secondary photoresponse at the metal–semiconductor interface is responsible for the measured high-gain effect.

## ACKNOWLEDGEMENT

This work was funded by the National Science Council of Taiwan (Grant Nos. NSC 101-2112-M-002-026). H.Y.L. acknowledges support by the Aim for Top University Project of National Taiwan University (Grant Nos. 102R4000). We would like to thank Prof. Chi-Te Liang for helpful discussions.

Table 1. Data extracted from I-V characteristics for the calculation of the secondary photoresponse model. All of the measured photocurrents were performed under 0.5 V reversed bias, and the wavelength of the illuminated light is $\lambda$ = 390 nm.

| Parameter | Dark | Under Illumination |
|---|---|---|
| Ideality factor ($n$) | 2.45 | 16.2 |
| Schottky Barrier Height (SBH) (eV) | 0.82 | 0.58 |
| Measured photocurrent (A) | $4.30 \times 10^{-5}$ | $2.98 \times 10^{-4}$ |

Device area $A = 4.30 \times 10^{-5}$ m$^2$

Temperature $T = 302$ K

Figure Captions.

Fig. 1  RT PL spectra of ZnO with $H_2O_2$ treatment of different durations, with two arrays indicating the broadened DL emission The table in the inset shows the type of defects and the corresponding wavelengths.

Fig. 2  Experimental fitted SBH versus $n$ plot for ZnO samples grown at 600 °C. The vertical dotted line indicates $n = 1$, which is a perfect diode. The inset shows the I-V characteristics of a selected ($n = 2.45$ and SBH = 1.42 eV) ITO/ZnO Schottky photodiode.

Fig. 3  Calculated $R$ of the photodiode from the measured photocurrent $I_p$. The inset picture shows the optical power of the Xe-lamp used to perform the measurement.

Figure 1

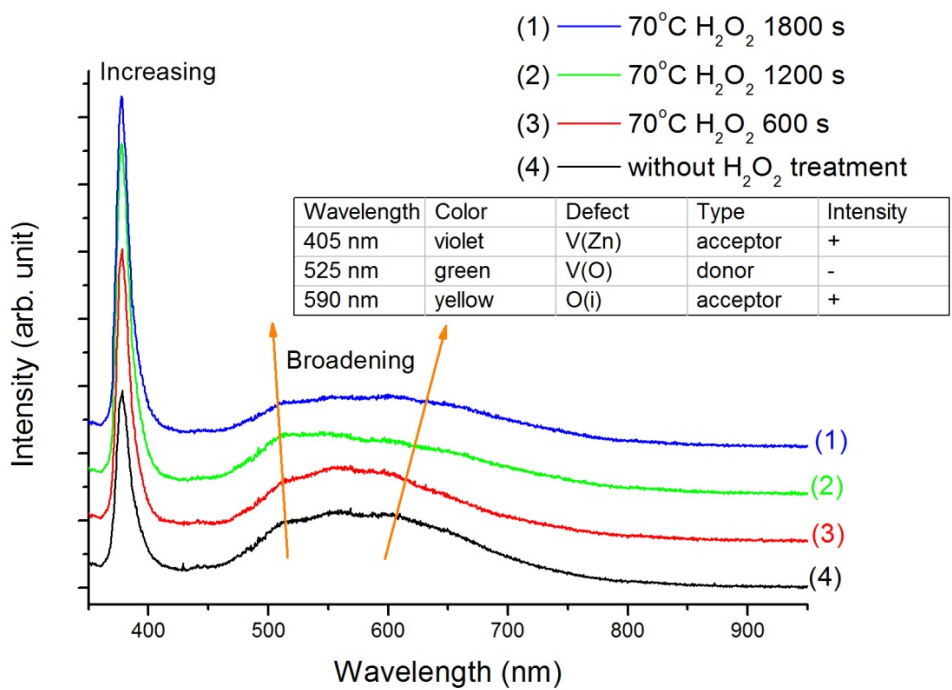

Figure 2

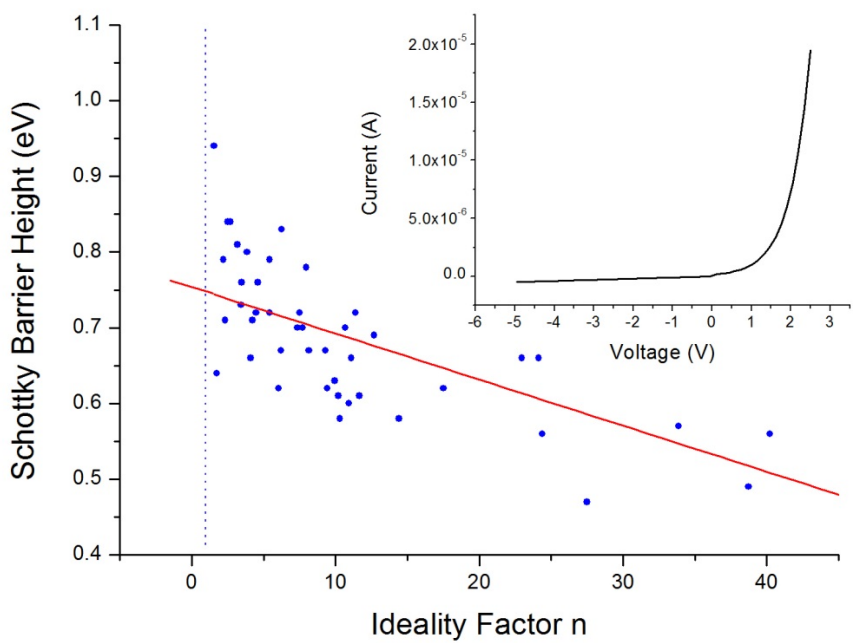

Figure 3

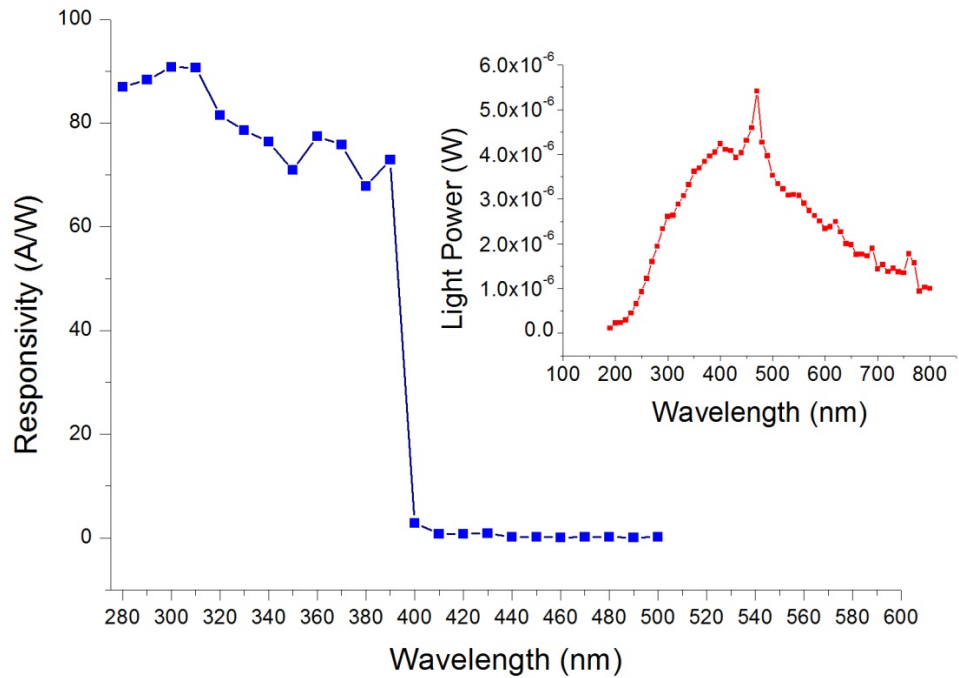